# The Feasibility of Coherent Energy Transfer in Microtubules

Travis John Adrian Craddock[1,2,3], Douglas Friesen[4], Jonathan Mane[4], Stuart Hameroff[5], and Jack A. Tuszynski[4,6]

[1]Center for Psychological Studies, Nova Southeastern University, Ft. Lauderdale, FL, USA, 33314; [2]Graduate School of Computer and Information Sciences, Nova Southeastern University, Ft. Lauderdale, FL, USA, 33314; [3]Institute for Neuro-Immune Medicine, Nova Southeastern University, Ft. Lauderdale, FL, USA, 33328; [4]Department of Oncology, University of Alberta, Cross Cancer Institute, Edmonton, AB, Canada, T6G 1Z2; [5]Departments of Anesthesiology and Psychology, Center for Consciousness Studies, The University of Arizona Health Sciences Center, Tucson, AZ, USA, 210202; Department of Physics, University of Alberta, Edmonton, AB, Canada, T6G 2E1

Communicated by Travis Craddock, Center for Psychological Studies, Graduate School of Computer and Information Sciences, Institute for Neuro-Immune Medicine, Nova Southeastern University, 3440 South University Drive, Fort Lauderdale, Florida, United States of America, 33328, +1-954-262-2868, tcraddock@nova.edu

## Abstract

It was once purported that biological systems were far too "warm and wet" to support quantum phenomena mainly due to thermal effects disrupting quantum coherence. However recent experimental results and theoretical analyses have shown that thermal energy may assist, rather than disrupt, quantum coherence, especially in the "dry" hydrophobic interiors of biomolecules. Specifically, evidence has been accumulating for the necessary involvement of quantum coherence and entanglement between uniquely arranged chromophores in light harvesting photosynthetic complexes. Amazingly, the 'tubulin' subunit proteins, which comprise microtubules, also possess a distinct architecture of chromophores, namely aromatic amino acids including tryptophan. The geometry and dipolar properties of these aromatics are similar to those found in photosynthetic units indicating that tubulin may support coherent energy transfer. Tubulin aggregated into microtubule geometric lattices may support such energy transfer, which could be of import for biological signaling and communication essential to living processes. Here we perform a computational investigation of energy transfer between chromophoric amino acids in tubulin via dipole excitations coupled to the surrounding thermal environment. We present the spatial structure and energetic properties of the tryptophan residues in the microtubule constituent protein tubulin. Plausibility arguments for the conditions favoring a quantum mechanism of signal propagation along a microtubule are provided. Overall we find that coherent energy transfer in tubulin and microtubules is biologically feasible.

Keywords: energy transfer | quantum biology | optical spectra | microtubule | structure-based simulation

## Introduction

Since Schrodinger asked "What is Life?" (1) a possible link between the "weird" world of quantum phenomena and biology has been hypothesized. It has been often argued that living systems are too "warm, and wet" to support quantum effects, however recent cutting edge research suggests this is not necessarily always the case. From quantum coherence in photosynthesis and magnetoreception in birds to quantum olfaction (2) and individual photon effects in vision (3), the field of quantum biology is leaping into the mainstream. This leads us to inquire, where else might the effects of quantum biology be found?

Without a doubt, the standard bearer for the emergence of quantum biology has been the coherence found in the light harvesting antennae utilized in photosynthesis. In 2007 Greg Engel et al. directly observed quantum oscillations caused by electronic coherence at 77 K in the Fenna-Matthews-Olsen (FMO) photosynthetic light harvesting complex (LHC), a temperature pushing the "warm" limit for quantum processes, catapulting this field into the spotlight (4). This limit was further pushed to an astounding 277 K, nearing physiological temperature, and effectively ruling out the "warm and wet" limit for quantum phenomena in biology (5). Furthermore, these coherent effects do no seem to be restricted to the FMO complex alone, and have been shown for LHCs in plants (LHCII) (6-8), bacteria (LH2) (9,10) and phycobiliproteins (11,12).

The theoretical basis of this coherent energy transfer is the photo-induced interaction of transition dipoles in chlorophyll molecules. The unique chromophoric nature of chlorophyll and the elegant geometrical arrangement of these pigments in the light harvesting complexes of plants and bacteria allow for photon energy to be efficiently captured and funneled from environment to reaction centers. This biological light-harvesting critically depends on the quantum mechanism through which light energy hops between chromophores (13). An initially excited donor chromophore can convey its electronic energy to an acceptor chromophore via electrodynamic coupling of their transition electric dipole moments due to the close correspondence, or 'resonance', between their energy levels. This resonance energy transfer (RET), which was original pioneered by Forster (14) and is more specifically termed FRET, remains the dominant theory applied in electronic energy transport (15,16). However, this process is not unique to photosynthesis. J-aggregates (a.k.a. Scheibe aggregates), collections of dye molecules, are an artificial form of a light-harvesting complex capable of capturing and manipulating photon energy (17,18). At the root, this phenomenon is attributable to the optimal packing of chromophores with a significant transition dipole.

Interestingly, tubulin, the microtubule constituent protein, possesses a network of chromophoric tryptophan amino acids (Fig. 1). The fluorescence quantum yield for pure tryptophan is 0.14, and the experimentally observed yield for wild-type tubulin is 0.06 at room temperature (19). These are comparable to the room temperature fluorescence quantum yields of bacteriochlorophyll of 0.18, and the yield for LH1 of 0.08, and LH2 of 0.10 (20), respectively. The 'red edge effect' has also been observed in tubulin (21) indicating Trp to Trp energy transfer (22). Thus, these may serve as potential conduction pathways in tubulin and microtubules (23). Could these aromatic chromophores support coherent quantum effects within tubulin? As persistent electronic coherences may be a general property of any system of compact nearly static chromophores coupled to the environment (24), we investigate the biological feasibility of coherent energy transfer in the tubulin dimer. Via a combination of homology modeling, molecular dynamics (MD) simulations, quantum chemistry and optical biophysics we apply structure-based simulations similar to current studies of the FMO complex (25,26), and larger LHCs (27) to probe potential energy transfer mechanisms in tubulin, and by extension microtubules.

## Results and Discussion

The dominant spatial distribution and orientation of the tryptophan residues (Trps) in tubulin after 8 ns of molecular dynamics simulation are shown in Fig. 1A. Between nearest-neighbors, spacing ranges between 11.4 and 41.6 Å (Fig. 1B). Importantly, this range is comparable to the distances between the bilin chromophores in cryptophyte marine algae, which has been shown to support quantum-coherent transfer of electronic excitation (11).

This spatial range, i.e. the average separation between neighboring dipoles, is considered larger than the extension of the Trp transition densities and so the Point-Dipole (PD) approximation was used to determine the inter-Trp excitonic couplings. In calculating the excitonic interaction coefficients, the choice of the dielectric constant value, and the method chosen for screening and local field effects is no trivial matter (28). Typically, effective protein dielectric constant values are chosen between 2 and 4, however experimental measures of the dielectric coefficients in protein hydrophobic pockets indicate this value is well above this range, arguing against this practice (29). The experimentally measured tubulin optical dielectric value of 8.41 falls well outside this range (30), and is expected to have a significant impact on coupling strengths.

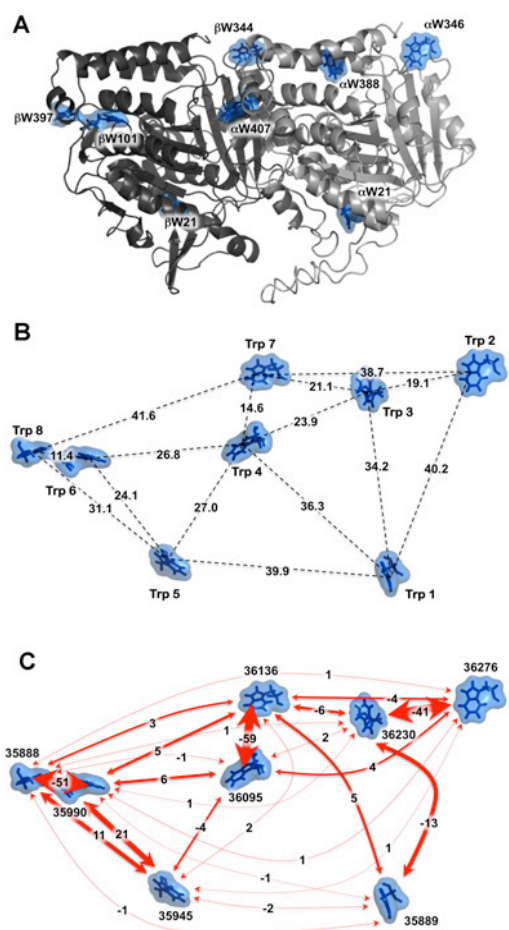

**Fig. 1:** Arrangement of eight Trp chromophores in one tubulin αβ-dimer. Laterally oriented with the microtubule protofilament axis along the horizontal, microtubule surface above, microtubule lumen below. C-terminal tails not shown. (A) Location within tubulin dimer. (B) Spacings in angstrom. (C) Dipole couplings in cm$^{-1}$ for a dielectric constant of 8.41.

In the Förster approximation, accounting only for electrostatic screening effects due to the porlarizable medium being present between the dipoles, the dipolar couplings are small indicating little to no interaction between Trps regardless of the dielectric value chosen.

**Table 1:** Site and Interaction Energies in Units of cm$^{-1}$ (Diagonal, shown in bold, are with respect to 35,688 cm$^{-1}$).

| Trp | 1 | 2 | 3 | 4 | 5 | 6 | 7 | 8 |
|-----|---|---|---|---|---|---|---|---|
| 1 | **+1** | 0 | -13 | 0 | -2 | -1 | 5 | -1 |
| 2 | 0 | **+388** | -41 | 4 | 1 | 1 | -4 | 1 |
| 3 | -13 | -41 | **+342** | 2 | 0 | 1 | -6 | 1 |
| 4 | 0 | 4 | 2 | **+207** | -4 | 6 | -59 | -1 |
| 5 | -2 | 1 | 0 | -4 | **+57** | 21 | 2 | 11 |
| 6 | -1 | 1 | 1 | 6 | 21 | **+102** | 5 | -51 |
| 7 | 5 | -4 | -6 | -59 | 2 | 5 | **+248** | 3 |
| 8 | -1 | 1 | 1 | -1 | 11 | -51 | 3 | **0** |

This is consistent with previous estimates of Förster dipole-dipole energy transfer between Trps in tubulin (19)[*]. However, accounting for electrostatic screening effects of the medium as well as local inhomogeneous field effects yields different results depending on the dielectric constant chosen. As the dielectric constant value increases between 2 and 8.41 the coupling strengths become increasingly non-negligible.

It must be noted here that using a dielectric constant value above 4 with the screening factor $f = \varepsilon_{opt}^{-1}[(\varepsilon_{opt}+2)/3]^2$ yields $f > 1$. While this seems somewhat paradoxical in terms of a screening factor enhancing coupling, we consider it in the quantum viewpoint of Knox and van Amerongen (28) as the effect of the protein environment on the Trp dipole strengths. For the experimentally measured optical dielectric constant, coupling strengths up to 60 cm$^{-1}$ are found with the most significant couplings being between Trps 2 and 3, Trps 4 and 7, and Trps 6 and 8. All coupling strengths for the dielectric constant of 8.41 are shown in Fig. 1C which accounts for the geometrical arrangement in place.

Calculated site energies using a dielectric constant value of 8.41 range between 35,888 and 36,276 cm$^{-1}$, giving the maximum energy difference between Trps of 388 cm$^{-1}$. These site energies are shown in Fig. 1C and Table 1. Diagonalization of the Hamiltonian matrix (Table 2) revealed that the degree of delocalization of excited state energies is relatively small. Only for the exciton states at 36,054, 36,176 and 36,300 cm$^{-1}$ is there a significant delocalization over more than one pigment, with meaningful contributions between Trps 4 and 7, and Trps 2 and 3, due to the close site energies between these pigments and their relatively large couplings.

While the coupling strengths become increasingly non-negligible as the dielectric constant value increases above 2, the general line shape of the simulated spectra does not align well with experiment for values of 2 and 4.

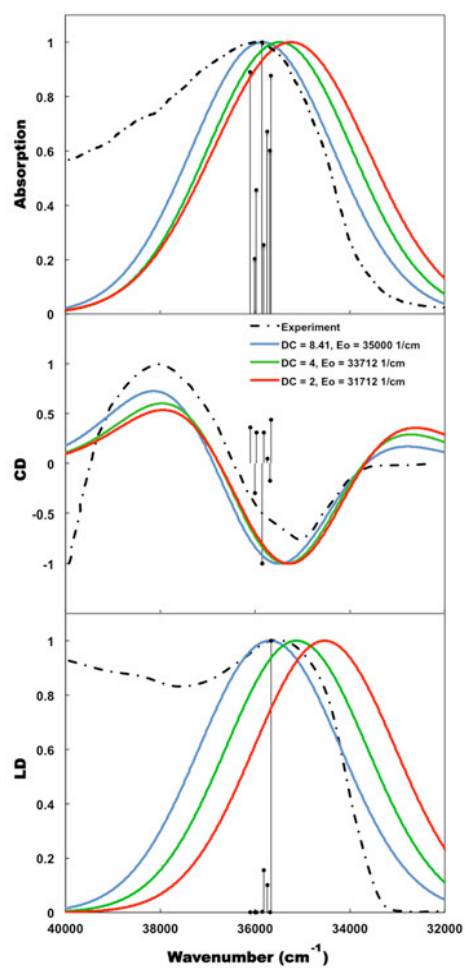

**Fig. 2:** Alignment of calculated spectra with experiment. Experimental spectra, bare calculated stick spectra for DC 8.41, and calculated spectra with Gaussian broadening applied for the Absorption, Circular Dichroism (CD) and Linear Dichroism (LD) of tubulin at 300 K. All values are plotted in arbitrary units similar to Vulto et al. (77).

This is most prominently seen in the CD spectra, and is not accounted for by simple shifts in $E_0$ alone (Fig. 2). However, when the experimentally measured optical dielectric value of tubulin is used the simulated spectra align well with experiment (see Fig. 2). This supports the use of the experimentally measured dielectric value over values between 2 and 4.

---

[*] This study also calculates negligible intramolecular Trp-Trp energy transfer efficiencies in tubulin, however these are based on an estimated dielectric constant of 2 and an assumed spectral overlap integral for NATA in THAM buffer at pH 7.8. As Trp is highly sensitive to its local environment (76) these results are seemingly inaccurate as the calculation does not accurately reflect the protein environment.

**Table 2:** Excited-State Energies (cm$^{-1}$), and Coefficients of the Eigenvectors Resulting from Diagonalization of the Hamiltonian in Table 1. Coefficients larger than 50% are shown in bold.

|          | Contribution of Trp no. | | | | | | | |
|----------|-------|-------|-------|-------|-------|-------|-------|-------|
| energies | 1     | 2     | 3     | 4     | 5     | 6     | 7     | 8     |
| 35663    | 0.04  | 0.00  | 0.00  | -0.02 | -0.22 | 0.40  | -0.02 | **0.89** |
| 35688    | **-1.00** | 0.00 | -0.04 | 0.01 | -0.05 | 0.01 | 0.02 | 0.02 |
| 35746    | -0.04 | 0.00  | 0.00  | 0.03  | **0.95** | -0.12 | 0.00 | 0.28 |
| 35814    | 0.01  | 0.00  | 0.00  | 0.12  | -0.22 | **-0.90** | 0.09 | 0.35 |
| 35854    | 0.02  | 0.00  | 0.01  | **0.81** | 0.01 | 0.15 | **0.57** | -0.04 |
| 35976    | 0.00  | -0.14 | -0.21 | **0.56** | -0.02 | 0.00 | **-0.79** | -0.01 |
| 36008    | 0.04  | -0.48 | **-0.84** | -0.14 | 0.00 | 0.00 | 0.21 | 0.00 |
| 36100    | -0.02 | **-0.86** | **0.50** | -0.01 | 0.00 | 0.00 | 0.00 | 0.00 |

Population dynamics calculations from the Haken-Strobl (31) method were applied to the Hamiltonian in Table 1 for initial excitations occurring on each of the 8 Trp residues (see Fig. 3). It was determined that coherent beatings last for ~600 fs with a pure dephasing rate of 50 cm$^{-1}$, when the excitation starts at Trps 2, 3, 4 or 7, and for < 300 fs when starting at Trp 6 or 8. Excitations starting at Trp 1 or 5 did not produce any beating effects. When starting at Trp 2 or 3 the excitation can be seen to spread to Trp 4 and 7 reaching close to the equal population distribution of 0.125 within ~2 ps. Negligible exciton populations in the remaining Trps are observed in this time frame. When starting at Trp 4 or 7, the same effect is observed for Trps 2 and 3 with additional increases in exciton population for Trps 5, 6, and 8 to ~0.06, spreading over the entire dimer.

Starting from the highest energy state the excitation travels from Trp 2 → 3→ 7 → 4 → 6 → 8 covering the length of the tubulin dimer. Stacking tublin dimers end to end in a microtubule would most likely alter this energy landscape, however it is feasible that such an arrangement could effectively transfer energy along the protofilament length as the inter-dimer spacing of Trps is uninterrupted (Fig. 4). Additionally, the unique cylindrical lattice symmetries found in the tubulin lattices of microtubules may effectively serve to enhance transfer rates and distances, and potentially enable energy transfer along helical pathways. Generalized Förster effects (32), induced by geometrical symmetries, can enhance the exciton diffusion length along cylindrically symmetric structures (33), similar to the helical arrangement of tubulin, and it's chromophores, in microtubules.

Is this effect unique to tubulin and microtubules? Trp is an ubiquitous amino acid found in most all proteins in low abundance, and FRET between Trp residues and other fluorophores is a technique commonly used to probe protein structure and dynamics. In most cases this RET is restricted to a single isolated protein, or between two adjacent interacting proteins or molecules. Crystalized globular proteins have the potential to efficiently transfer energy between Trp residues (34), yet these are rarely, if ever, found in biology. Protein polymers, such as the cell cytoskeleton, on the other hand are pervasive in biology, offering candidate structures for this type of energy transmission. While crystal structures for the majority of cytoskeletal components do not currently exists, comparison of their amino acid sequences reveals that the majority of intermediate filaments have a Trp abundance less than a third of that found in tubulin (Table 3) suggesting unfavorable conditions for RET. Only actin and synemin possess comparable densities, suggesting that polymers of these molecules could support RET similar to that in tubulin and microtubules. Although the crystal structure of actin is available and reveals 4 Trp per subunit with intramolecular separations ranging between 10 and 23 Å, the intermolecular distances between these Trp clusters is > 40 Å (Fig. 4) suggesting unfavorable conditions for RET within a microfilament.

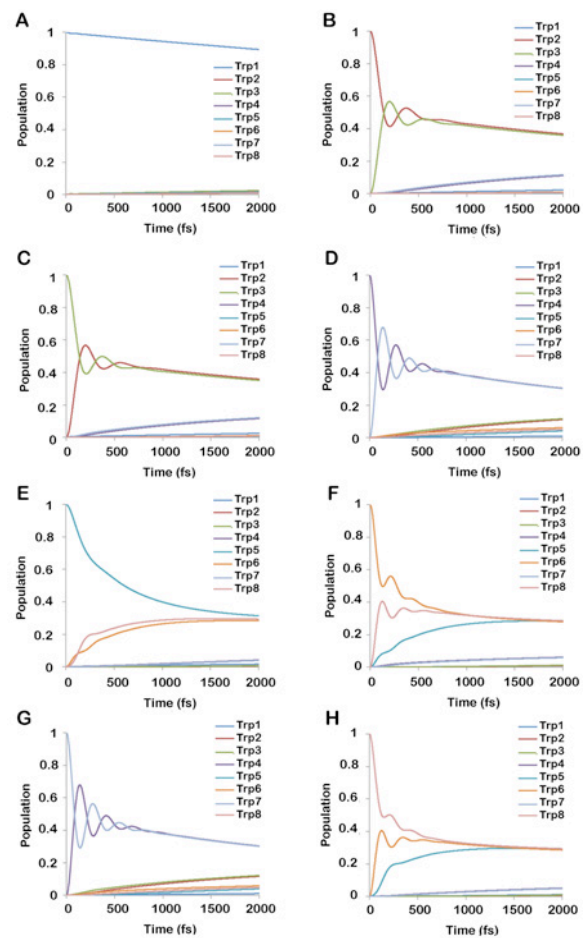

**Fig. 3:** Time evolution of the exciton population of each tryptophan in tubulin via the Haken-Strobl-Reineker model. Initial pure states are set as (A) Trp1: αW21, (B) Trp2: αW346, (C) Trp3: αW388, (D) Trp4: αW407, (E) Trp5: βW21, (F) Trp6: βW101, (G) Trp7: βW344 and (H) Trp8: βW397.

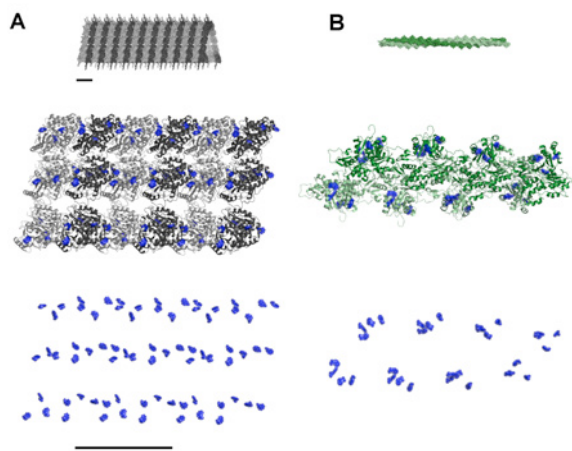

**Fig. 4:** Tryptophan architectures in cytoskeletal filaments. (A) Tubulin based microtubule. (B) Actin based microfilament. Scale bars are approximately 10 nm. Actin microfilaments are built using PDB ID: 2ZWH using the Oda et al. filament model (78).

Whether this has a biological role remains open. Such rapid signaling through Trp conduction pathways may coordinate the complex organization of the microtubule cytoskeleton required for the tasks of cell division, motor protein trafficking, and motility. Microtubules have been shown to reorganize in a dose-dependent manner after exposure to UV light (35-38), with the greatest effect being observed around 280 nm (38). Feasible mechanisms for these changes include the reduction of disulfide or peptide bonds induced by photoexcitation of tryptophan groups (39-42), or subtle protein structural changes due to photoinduced alterations in Trp flexibility (42). Such a signaling mechanism may explain observed the apparent UV mediated cell-to-cell influence on cell division (43). Only further investigation, both theoretically and experimentally, will tell.

Thus, we conclude that, based on this initial analysis, the unique Trp network within an individual tubulin dimer can possess significant dipolar couplings capable of supporting quantum coherent beating effects similar to those observed in the FMO photosynthetic complex (4,5), LHCII (7,8), LH2 (9,10) and phycobiliprotein LHCs (11,12). Furthermore, our results suggest that this network may support coherent energy transfer at physiological temperature between clusters of Trps in tubulin, and microtubule structures. One of the main objectives of this paper is to stimulate future experimental work in this area. While this is purely a computational prediction it supports the continued investigation of coherent energy transfer mechanisms in tubulin and microtubule lattices. Our predictions are experimentally feasible to verify employing the same methods as those used in the case of photosynthetic complexes.

**Table 3:** Tryptophan abundance in cytoskeletal constituent molecules

| Cytoskeletal Molecule | Total Residues | Trps | Trp Abundance |
|---|---|---|---|
| Tubulin monomer | ~450 | 4 | 0.9% |
| Actin | 375 | 4 | 1.1% |
| α-Internexin | 499 | 1 | 0.2% |
| Desmin | 470 | 1 | 0.2% |
| GFAP | 432 | 1 | 0.2% |
| Keratin | 564 | 2 | 0.3% |
| Neurofilament | 916 | 3 | 0.3% |
| Peripherin | 470 | 1 | 0.2% |
| Syncoilin | 482 | 0 | 0.0% |
| Synemin | 1565 | 15 | 1.0% |
| Vimentin | 466 | 1 | 0.2% |

## Methods

Here, we give a brief summary of the computational procedures yielding site energies, excitonic couplings, optical spectra, and time evolution of exciton populations. All data will be made available on request.

**Molecular Dynamics Simulation/Dominant Tryptophan Confirmation:** MD Simulations were performed based on previous methods (44,45), using NAMD (46). Coordinates for the αβ-tubulin dimer at 3.5 Å, GDP, GTP, and magnesium ion are from the Brookhaven National Laboratory Protein Data Bank (47) entry 1JFF (48), and PDB2PQR 1.7 (49) was used to generate per-atom charge and radius, at pH 7.0 using PROPKA (50,51), and the AMBER (52) force field. Swiss-PdbViewer 4.0 (53) was used to model the missing residues in the crystal structure (α-tubulin: 1, 35-60, 440-451, β-tubulin: 1, 438-455). GDP and GTP were modelled for use by the AMBER 94/99 force field by Meagher et al. (54). PTRAJ, using the AMBER99SB force field (55) from AMBER 10 (52) was used to neutralize the system, add counterions to reproduce physiological ionic concentrations, and add a TIP3P water cube buffer of 25 Å (63283 molecules of water). After minimization and heating to 310 K, equilibration occurred for 519 picoseconds (ps) while gradually releasing constraints on backbone atoms. Then, using periodic boundary conditions, a 17 nanosecond (ns) simulation occurred where atomic co-ordinates were saved from the trajectory every 2 ps.

PTRAJ was used to analyze the 17 ns trajectory for root mean square deviation (RMSD) of the backbone of non c-terminus residues of both alpha and beta tubulin, as well as the beta-factor of all residues. The RMSD was found to stabilize after approximately 8 ns. The final 8 ns of the total trajectory was chosen for further analysis. The GROMACS program g_cluster (56) was used to cluster the Trp atom positions of all frames via single linkage with a cutoff of 0.5 Å. The middle structure of the largest cluster was taken as the dominant conformation for the Trp residues.

**Vertical excitation energies for Tryptophan /Site Energy Calculations**: Site energies were derived from structure-based calculations of the free energy change of the protein-pigment complex (PPC) that occurs when the ground-state charge density of Trp m is shifted to the first excited state, analogous to methods used for the FMO complex (26,57,58).

Quantum chemical calculations of the pigments *in vacuo*, yield the charge distributions of the $S_0$ and $1L_a$ states and a contribution $\Delta E_{qm}$ to the $S_0 \rightarrow 1L_a$ transition energy due to the differing orientations of the Trps. Trp side chains were each isolated from the dominant conformation (described above) and capped with a hydrogen atom to complete the valence. *Ab initio* geometry optimizations at the ground state were performed using density functional theory (DFT) at the B3LYP/6-311G(d,p) (59) level. Quantum chemical calculations of the 8 lowest singly excited vertical excitation energies of Trp were done using time-dependent density functional theory (TDDFT) (60,61) at the B3LYP/6-311+G(d,p) (59) level. All calculations were performed using ORCA v2.8 (http://cec.mpg.de/forum/).

The vertical transition energies for the $1L_a$ state of the Trp side chains were found to be comparable to previous work using similar methods (62), however like these results our values were found to be lower than experimentally determined values (62,63). To align with experiment (38473 cm$^{-1}$) a scaling factor of 1.02 was applied to the calculated values. The relative quantum correction, $\Delta E_{qm}$, to the site energy was taken as the difference between the scaled and the experimental values.

The second part of the site energy calculation results in a contribution $\Delta E_{coul}$ due to classical electrostatic interaction between the charge distributions of the $S_0$ and $1L_a$ states with the protein environment. Using the Charge Density Coupling (CDC) (57,58) method the site energy shift of the m$^{th}$ Trp is calculated from the Coulomb interaction of the difference of the $S_0$ and $1L_a$ state partial charges $\Delta q_i^{(m)}$, obtained from the Löwdin atomic point charges in the above quantum chemical calculations, with the remaining background charges of the protein $q_j^{(bg)}$. As there can be large uncertainties when representing the Trp ring charge distribution as Löwdin atomic point charges, we have chosen to empirically scale the charges by a factor of 0.80 to improve alignment with the experimental spectra (described below). The electrochromic shift $\Delta E_{coul}$ is given as,

$$\Delta E_{coul} = \frac{1}{\varepsilon} \sum_{i=1}^{N} \sum_{j=1}^{K} \frac{\Delta q_i^{(m)} \cdot q_j^{(bg)}}{\left| r_i^{(m)} - r_j^{(bg)} \right|}$$

where N is the number of partial charges of the m$^{th}$ Trp side chain, K is the total number of background partial charges (including other Trp residues), and $|r_i^{(m)}-r_j^{(bg)}|$ is the distance between the i$^{th}$ difference partial charge of the m$^{th}$ pigment and the j$^{th}$ partial charge of the background. Background charges were taken from the AMBER99SB force field (55). Here we use the measured high frequency dielectric for tubulin setting $\varepsilon = 8.41$ (30).

The site energy for the m$^{th}$ Trp is thus given as,

$$E_m = E_0 + \Delta E_{qm} + \Delta E_{coul}$$

where $E_0$ is taken as 35000 cm$^{-1}$ (285 nm) to align with experimental spectra (described below).

**Excitonic Coupling Interactions**: Excitonic couplings are calculated using the Point-Dipole (PD) Approximation, valid as the distance between pigments is large compared to the extension of their transition densities. The PD excitonic coupling is given as,

$$V_{mn} = f \frac{\mu_{vac}^2}{R_{mn}^3} \left[ \vec{e}_m \cdot \vec{e}_n - 3(\vec{e}_m \cdot \vec{e}_{mn})(\vec{e}_n \cdot \vec{e}_{mn}) \right]$$

where $\vec{e}_m$ is a unit vector along the transition dipole moment of the m$^{th}$ Trp, the unit vector $\vec{e}_{mn}$ is oriented along the line connecting the centers of Trps m an n, $\mu_{vac}$ is the transition dipole moment of the $1L_a$ transition of Trp in vacuum, and the factor f effectively takes into account local field and screening effects in an effective way. In the simplest approximation due to Förster, $f = 1/\varepsilon_{opt}$, accounting only for screening effects. However, when screening effects of the medium as well as local field effects are included, this becomes $f = \varepsilon_{opt}^{-1}[(\varepsilon_{opt}+2)/3]^2$ (64-67). Here we consider both these cases. The optical dielectric constant for a typical protein environment is given in the range of 2 – 4, however, tubulin's high frequency dielectric coefficient has been measured as 8.41 (30). For comparison, we consider the two extremes of 2 and 8.41.

**Alignment with experimental spectra**: The tight-binding Hamiltonian for an interacting N-body system in the presence of a single excitation is given by (68),

$$H_S = \sum_{m=1}^{N} \varepsilon_m |m\rangle\langle m| + \sum_{m<n}^{N} V_{mn}(|m\rangle\langle n| + |n\rangle\langle m|)$$

where the states $|m\rangle$ denote the excitation being at site m. The site energies and coupling terms are given by $\varepsilon_m$ and $V_{mn}$, respectively, and are calculated as described above. Exciton stick spectra for absorption, circular dichroism (CD) and linear dichroism (LD) were calculated from the orientations of the Trp molecules in the dominant confirmation and the diagonalized Hamiltonian matrix using the formula described by Pearlstein (69). Experimental spectra were calculated by applying a Gaussian function to each exciton transition using an assumed full-width half maximum value of 3,500 cm$^{-1}$ as commonly observed for Trp (70,71). These were then compared to experimental absorption, CD, and LD spectra adapted from Mozzo-Villarías et al. (72), Clark et al. (73), and Marrington et al. (74), respectively.

**Exciton Population Dynamics:** Time evolution of the exciton population in the Trp network in the presence of thermal fluctuations of the environment is modeled using the Haken and Strobl model (31). Here it is assumed that thermal fluctuations of the environment couple to the Trps by an electron-phonon interaction, acting only on the diagonal elements of the Hamilitonian system $H_S$, with the fluctuations being unbiased, uncorrelated and Gaussian in nature. Under these assumptions the Haken-Strobl equation for the density operator ρ in the Schrödinger picture is given as,

$$\dot{\rho}(t) = -\frac{i}{\hbar}[H_S, \rho(t)] + L_\phi(\rho(t))$$

where the pure-dephasing Lindblad operator is given by,

$$L_\phi(\rho(t)) = \sum_{m=1}^{N} \gamma_m \left[ A_m \rho(t) A_m^\dagger - \frac{1}{2} A_m A_m^\dagger \rho(t) - \frac{1}{2} \rho(t) A_m A_m^\dagger \right]$$

with $A_m$ and $A_m^\dagger$ being the creation and annihilation operators, and $\gamma_m$ the pure dephasing rates for the m$^{th}$ Trp. As a reasonable assumption here we take $\gamma_m$ to be 50 cm$^{-1}$ for all sites. All population dynamics were computed using QuTip 2 (75).

## Acknowledgements

Thanks to Dr. Chih-Yuan Tseng at Sinoveda Canada Inc. for his assistance in repairing the missing residues of the 1JFF crystal structure of the tubulin dimer, Dr. J. Robert Johansson at the RIKEN, Advanced Science Institute for his help with the QuTip programming package, and Dr. David Sept at the University of Michigan for supplying files for the actin and microfilament structures. This research has been enabled by the use of computing resources provided by WestGrid and Compute/Calcul Canada.